\documentclass[twocolumn,showpacs]{revtex4}
\usepackage{tabularx}
\usepackage{dcolumn}
\usepackage{graphics}
\usepackage{bm}
\usepackage[dvips]{graphicx}
\usepackage[dvips]{rotating}
\usepackage[latin1]{inputenc}
\usepackage{dcolumn}
\usepackage{tabularx}
\usepackage{amsmath}
\usepackage{amssymb}
\usepackage{float}
\begin{document}
\draft

\title{Field-ionization threshold and its induced ionization-window phenomenon for Rydberg atoms in a short single-cycle pulse}

\author{B. C. Yang and F. Robicheaux}
\email{robichf@purdue.edu}

\affiliation{Department of Physics, Purdue University, West
Lafayette, Indiana 47907, USA}

\date{\today}

\begin{abstract}
We study the field-ionization threshold behavior when a Rydberg atom
is ionized by a short single-cycle pulse field. Both hydrogen and
sodium atoms are considered. The required threshold field amplitude
is found to scale \emph{inversely} with the binding energy when the
pulse duration becomes shorter than the classical Rydberg period,
and, thus, more weakly bound electrons require larger fields for
ionization. This threshold scaling behavior is confirmed by both 3D
classical trajectory Monte Carlo simulations and numerically solving
the time-dependent Schr\"{o}dinger equation. More surprisingly, the
same scaling behavior in the short pulse limit is also followed by
the ionization thresholds for much lower bound states, including the
hydrogen ground state. An empirical formula is obtained from a
simple model, and the dominant ionization mechanism is identified as
a nonzero spatial displacement of the electron. This displacement
ionization should be another important mechanism beyond the
tunneling ionization and the multiphoton ionization.  In addition,
an ``ionization window'' is shown to exist for the ionization of
Rydberg states, which may have potential applications to selectively
modify and control the Rydberg-state population of atoms and
molecules.

\par


\end{abstract}

\maketitle

Studies on the ionization of atoms and molecules in an external
field(s) have greatly broadened our knowledge of the microscopic
electron dynamics, and also deepened our understanding on the
correspondence between quantum and classical
mechanics\cite{external_field, intense_field, Gallagher,
Gutzwiller}. Correspondingly, various ionization mechanisms have
been identified for a diversity of interesting phenomena in
different field configurations, such as closed-orbit
theory\cite{COT}, ``simple-man's'' model\cite{simple_man00,
simple_man01, simple_man02}, energy-level splitting and
crossings\cite{ramp}, successive Landau-Zener
transition\cite{micro_wave}, dynamical
localization\cite{localization}, and also impulsive-kick ionization
in a short half-cycle pulse (HCP) \cite{half00, half01, half_add}.
Recently, an intense single-cycle THz pulse has been applied to
explore the ionization dynamics for low-lying Rydberg states of
sodium atoms\cite{Bob}, where the field ionization threshold was
found to scale as $n^{-3}$ ( $n$ denotes the principal quantum
number), in contrast with all the threshold behaviors discovered
before\cite{ramp, micro_wave, localization, half00, half01}. The
threshold value is defined as the required field amplitude
$F_{10\%}$ for $10\%$ ionization probability. Different threshold
behavior corresponds to different ionization mechanism. This novel
threshold behavior indicates that, when the pulse duration $t_w$
becomes comparable with (or even smaller than) the Rydberg period
$T_{Ryd}=2\pi n^3$ (atomic units are used unless specified
otherwise), the possible time effects imprinted by a short
single-cycle pulse can be expected in the ionization dynamics.

In this paper, a new counterintuitive threshold scaling behavior is
found in the short single-cycle pulse limit for both Rydberg states
and much lower bound states, including the hydrogen ground state.
The required threshold field amplitude is proportional to
$(n/t_w)^2$, which suggests that a stronger threshold field is
required for higher Rydberg states. This threshold behavior is
confirmed by comparing 3D classical trajectory Monte Carlo (CTMC)
simulations with numerical results from directly solving the
time-dependent Shr\"{o}dinger (TDS) equation. It is further
supported by a simple model where the dominant ionization mechanism
is identified as a sudden and finite displacement of the electron in
the short pulse duration. This displacement-ionization mechanism
adds a new element in the strong-field ionization regime, which has
received little attention before.

By combining with the adiabatic-ionization threshold in the
low-frequency limit\cite{ramp}, it can be shown that there is an
``ionization window'' in the Rydberg series. The location of this
``ionization window'' can be adjusted by the pulse duration, and its
width and height are dependent on the field strength. Since the
generation of Rydberg states has been a routine, and the
selective-field-ionization (SFI) technique has also been developed
to identify the Rydberg-state population in atoms and
molecules\cite{Gallagher}, this ``ionization window'' phenomenon may
have potential applications in future experiments to modify and
control the Rydberg-state population.

\begin{figure*}[t]
\centering
  \includegraphics[width=500pt]{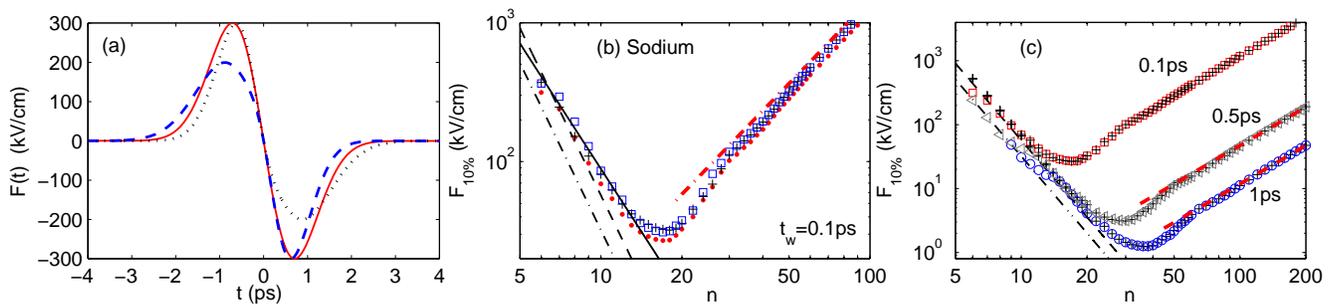}
  \caption{(Color online) (a) presents three typical single-cycle pulses.
  The solid, the dashed and the dotted curves represent, respectively, a ``symmetric'',
  an ``asymmetric'' and an ``inverted'' field profiles given by Eq.(1).
  (b) displays the threshold behaviors for sodium with $t_w=0.1ps$.
  The red solid points, the open squares and the points indicated by ``$+$''
  correspond respectively to a ``symmetric'' , an ``asymmetric'' and an
  ``inverted'' pulses. The bold dot-dashed line is from Eq.(4).
  The solid line indicates the $n^{-3}$ scaling behavior.
  (c) shows the threshold behaviors for both hydrogen and sodium
  with different $t_w$ values, where a ``symmetric'' pulse
  is applied. The blue circles, the gray triangles and
  the red squares display the thresholds for sodium
  with $t_w=1ps$, $t_w=0.5ps$ and $t_w=0.1ps$, respectively.
  The corresponding thresholds for hydrogen are denoted by
  ``$+$''. The bold dashed lines are from Eq.(5).
  In both (b) and (c), the thin dashed and the thin dot-dashed lines
  (black online) represent the thresholds for hydrogen and sodium in the long pulse
limit , respectively.}
\end{figure*}

The single-cycle pulses in our calculations are constructed by the
following vector potential,
\begin{equation}\label{1}
    A(t)=-\frac{F_{m}t_{w}}{C_0}e^{-\big[1\pm\frac{1}{a}\tanh\big(\frac{bt}{t_w}\big)\big]\big(\frac{t}{t_w}\big)^2},
\end{equation}
where $F_{m}$ and $t_w$ are, respectively, the pulse maximum
amplitude and the defined pulse duration; $a$ and $b$ are two
adjustable parameters, and $C_0$ is dependent on the values of $a$
and $b$. Here, we first set $a=3$, $b=2$, and $C_0=1.016$. The
signs``$+$'' and ``$-$'' in the exponential factor correspond to an
``asymmetric'' pulse as in Ref.\cite{Bob} and the ``inverted'' one,
respectively. An amplitude-symmetric pulse is given by Eq.(1) with
$b=0$ and $C_0=\sqrt{2}~e^{-1/2}$. Three typical pulses are plotted
in Fig.1(a) with $t_w=1ps$. The total pulse length is about $4t_w$.
For convenience, a related variable $\omega=\pi/(2t_w)$ is defined
to approximately denote the ``field frequency''\cite{FT}. The
applied pulse field ($F(t)=-dA(t)/dt$) is assumed to be linearly
polarized along the $z$-axis. A pure Coulomb potential is adopted
for hydrogen. For sodium atoms, the following model potential is
used,
\begin{equation}\label{2}
    V_m(r)=-\frac{Z^*(r)}{r}-\frac{\alpha}{2r^4}\Big(1-e^{-(\frac{r}{r_c})^3}\Big)^2
\end{equation}
where $r$ is the radial coordinate of electron relative to the
nucleus. $\alpha=0.9457$ and $r_c=0.7$.
$Z^*(r)=1+10e^{-a_1r}+a_2re^{-a_3r}$ with $a_1=3.8538$,
$a_2=11.0018$ and $a_3=3.0608$. Using 3D CTMC simulations\cite{MC00,
Topcu}, the calculated thresholds $F_{10\%}$ (field amplitudes $F_m$
for $10\%$ ionization probability) are displayed in Fig.1(b)-(c) as
a function of $n$, where $10^5$ trajectories are launched and the
Rydberg electron is assumed to be initially in a $d$ state with the
quantum angular momentum ($l=2$, $m=0$). The semiclassical angular
momentum $l+0.5$ is used in the classical simulations\cite{MC00}.

By applying the different pulses as those shown in Fig.1(a), the
thresholds for sodium are displayed, respectively, in Fig.1(b) with
$t_w=0.1ps$, which shows no qualitative change induced by the
detailed pulse shape. The other similar results are presented in
Fig.2, including those calculations with different initial angular
momenta. We find that the threshold behaviors are not sensitive to
the $l$ values, except that an effective quantum number $n^*$ should
be adopted for sodium by considering the larger quantum defect for
an $s$ state or a $p$ state induced by the ionic-core electrons.
Therefore, the results presented in the following discussions are
mainly for the atoms initially in a $d$ state, and a ``symmetric''
pulse is applied.

\begin{figure}[b]
\centering
  \includegraphics[width=220pt]{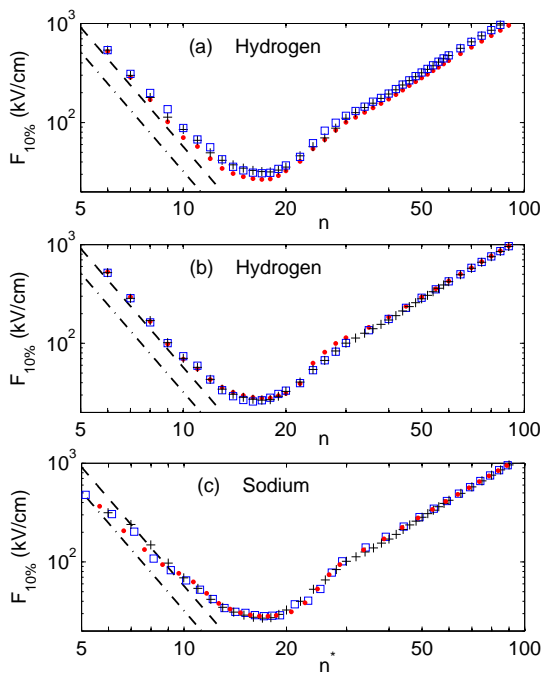}
  \caption{(Color online) Other similar results as those in Fig.1(b).
  (a) displays the threshold behaviors for hydrogen with different pulse profiles.
  The red solid points, the open squares and the points indicated by ``$+$''
  correspond respectively to a ``symmetric'' , an ``asymmetric'' and an ``inverted'' pulses.
  (b) shows the threshold values for hydrogen Rydberg states with different angular-momentum quantum
  numbers, where a ``symmetric'' field profile is used. The red solid points, the open squares and the points labeled by ``$+$''
  correspond respectively to the required threshold amplitudes for the Rydberg atom initially at an $s$ state, a $p$ state and a $d$ state.
  (c) is the same as (b) but for sodium, where an effective quantum number $n^*$ is used
  by including the quantum defects. In all the subplots (a)-(c), $t_w=0.1ps$.
  The dashed and the dot-dashed lines
  (black online) represent the thresholds for hydrogen and sodium in the long pulse
  limit , respectively. }
\end{figure}

For different field durations, a very similar threshold behavior is
observed except a global shift between the different cases (see
Fig.1(c)). This is not a surprise, because the classical Hamiltonian
is invariant by defining two scaled variables ($Fn^4$, $t_w/n^{3}$)
instead of the three quantities ($n$, $F$, $t_w$). As a result of
this invariance\cite{scaling, note}, all the threshold curves in
Fig.1 can be represented by one curve in the scaled space ($Fn^4$,
$\omega n^3$) in Fig.3(a), where the displayed points are directly
from the data in Fig.1(c) with $t_w=1ps$, and the results from other
$t_w$ values are not displayed due to indistinguishability.

In the regime near the thin dashed lines in Fig.1 and Fig.3(a),
where the pulse duration is much longer than $T_{Ryd}$, the
threshold behaviors can be understood based on a picture of the
energy-level splitting and crossings\cite{ramp}. For much lower
Rydberg states, the threshold for sodium is approximately equal to
$n^{-4}/16$ indicated by the dot-dashed line, in contrast with that
for hydrogen ($n^{-4}/9$) denoted by the thin dashed line. The
corresponding ionization mechanism has been identified as an
adiabatic over-the-barrier ionization, and the lower threshold for
sodium is a signature of an avoided-crossing effect between energy
levels due to the presence of ionic-core electrons. This
adiabatic-ionization threshold supplies a fundamental principle for
the SFI technique\cite{Gallagher, SFI}. With the field varying
faster, the threshold for sodium gradually deviates from $n^{-4}/16$
as a result of diabatic transition near the avoided crossings, and
finally behaves in the same way as that for hydrogen. This effect
can be observed clearly by comparing the threshold curves with
$t_w=0.1ps$ and $t_w=1ps$ in Fig.1(c).

When the pulse duration becomes comparable with $T_{Ryd}$, the
required threshold field strength can again deviate from the
diabatic ionization threshold, and the time effect seems to be
dominant in the ionization dynamics. For the initial stage, it has
been observed to scale as $n^{-3}$ recently\cite{Bob}, as that
indicated by the thin solid line in Fig.1(b). However, by going to
the higher-lying Rydberg states where the pulse duration is shorter
than $T_{Ryd}$, the required threshold will not continue to
decrease. We find that a larger field amplitude is required for a
more weakly-bound state. In the short pulse limit, the required
threshold is proportional to $n^2$ as shown by the bold dashed line
(red online) in Fig.1(c). Here, we would like to stress that, in the
short pulse regime, there is no qualitative discrepancy between the
threshold behaviors for sodium and hydrogen atoms.

To further confirm this counterintuitive behavior, a quantum
calculation is made for hydrogen $15d$ state at different ``scaled
frequencies'' ($\omega n^3$) by numerically solving the TDS
equation\cite{Topcu, TDS}. We represent the wave function on a 2D
space spanned by discrete radial points and an angular momentum
basis, where a split operator method is used with a Crank-Nicolson
approximation to propagate the wave function. For the radial part, a
square-root-mesh scheme is used with a Numerov approximation. The
results are displayed by the open squares in Fig.3(a), which agrees
with the classical simulations.

\begin{figure}[t]
\centering
  \includegraphics[width=220pt]{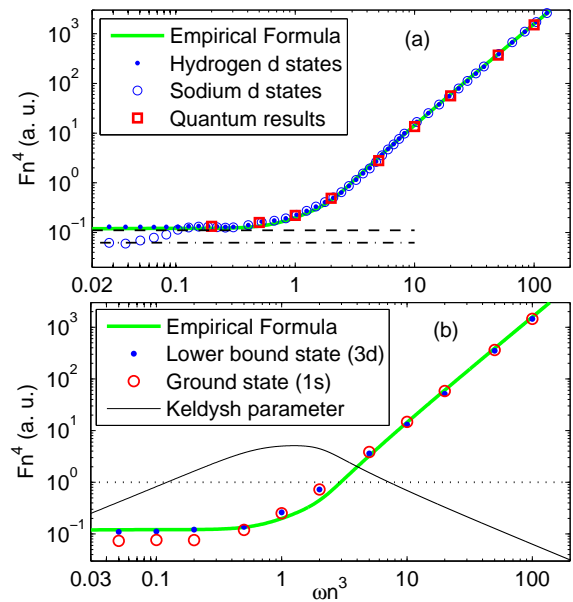}
  \caption{(Color online) Threshold scaling behaviors with (a) for Rydberg atoms and (b) for much lower bound states.
  $\omega$ is defined as $\pi/(2t_w)$. A ``symmetric'' pulse is
  applied. In (a), the open and the solid circles are from Fig.1(c) for $t_w=1ps$.
  The quantum results for hydrogen $15d$ state at different scaled frequencies are shown by the red open squares.
  The dashed and the dot-dashed lines represent the thresholds for hydrogen and sodium in the
  low-frequency limit, respectively. In (b), the open and the solid circles correspond to the thresholds for
  hydrogen atoms initially in $1s$ state and $3d$ state, respectively.
  In both (a) and (b), the bold solid curve (green online) is from Eq.(6). The thin solid
  curve in (b) shows the variation of Keldysh parameter $\gamma$ associated
  with the bold solid line. The unity value of $\gamma$ is indicated
  by the dotted line.}
\end{figure}

In the short pulse limit, the observed threshold can be understood
from a simple model. A similar idea was discussed in
Ref.\cite{displacement}. Consider a semiclassical atom with a
quantized total energy $E_n=-\frac{1}{2n^2}$, an electron moves
around the nucleus with $r_n=n^2$. Such a simple Bohr's model is not
fully correct for highly elliptical states, but, as a first
approximation, it provides a reasonable estimation and also insight
for the electron dynamics in an atom\cite{book}. By interacting with
a very short pulse field, the ionization can only occur during a
very short fraction of one Rydberg period. The electron experiences
a sudden displacement $\Delta r$, and its kinetic energy is assumed
to be unchanged approximately during the short-time displacement.
Following this simple assumption, the condition for ionization
should be
\begin{equation}\label{3}
    E_n+\frac{1}{r_n}-\frac{1}{r_n+\Delta r}\geq0~,
\end{equation}
which requires $\Delta r\geq n^2$. In a single-cycle pulse given by
Eq.(1), it can be shown that a freely-motion electron gets no
momentum transfer from the field, but has a finite displacement. For
a ``symmetric'' pulse, $\Delta r=\sqrt{(\pi e)/2}\cdot F_{m} t_w^2$.
Hence, the threshold field amplitude $F_{th}$ is
\begin{equation}\label{4}
    F_{th}=\sqrt{\frac{2}{\pi e}}\cdot\frac{n^2}{t_w^2}~.
\end{equation}
Note the symbol `$e$' on the right-hand side of Eq.(4) refers to the
base of natural logarithms. The predicted threshold in Eq.(4) is
displayed by the bold dot-dashed line in Fig.1(b), where a good
agreement with the numerical results can be observed except for a
small quantitative discrepancy. We attribute this small discrepancy
to the rough approximations in estimating $r_n$ and $\Delta r$.

By slightly modifying the constant coefficient on the right-hand
side of Eq.(4), a best fit can be obtained with the numerical
results, and we arrive at
\begin{equation}\label{5}
    F_{th}=\frac{\pi}{8}\cdot\frac{n^2}{t_w^2}
\end{equation}
which is shown by the bold dashed lines in Fig.1(c). By
incorporating the experimental observation\cite{Bob}
($F_{10\%}\varpropto n^{-3}$) in the middle regime as that shown in
Fig.1(b), we find that the threshold behavior can be described very
well by the following \emph{empirical} formula in the whole range
from a long pulse limit to a short pulse limit,
\begin{equation}\label{6}
    Fn^4=\frac{1}{9}e^{-\Omega}+\frac{1}{10}{\Omega}^{\frac{1}{3}}e^{-({\Omega}-1)^2}+\frac{1}{2\pi}{\Omega}^2e^{-(1/{\Omega})}
\end{equation}
where $\Omega=T_{Ryd}/4t_w$ ($=\omega n^3$) which can be considered
as an approximate ``scaled frequency''. The above \emph{empirical}
expression in Eq.(6) is a simple combination of the different
threshold behaviors in the three scaled-frequency regimes (an
exponential factor is used in each term on the right-hand side of
Eq.(6) to turn on the corresponding threshold scaling relation in
each regime): the first term corresponds to the hydrogen threshold
($F_{10\%}=1/(9n^4)$) in the low-frequency limit\cite{ramp}; the
second term is from the recent observation\cite{Bob}; and the last
term comes from Eq.(5) directly, corresponding to the threshold
behavior in the short pulse limit. The predicted threshold curves
are shown by the bold solid lines in Fig.3.

The same threshold scaling behavior in the short pulse limit is also
observed for much lower bound states. For hydrogen $3d$ and $1s$
states, respectively, the required thresholds at different scaled
frequencies are displayed in Fig.3(b) from numerically solving the
TDS equation. To estimate the Keldysh parameter $\gamma$ ($=\omega
n^3/(Fn^4)$) in Fig.3(b), Eq.(6) is used. The $\gamma$ value is
often used to estimate the importance of tunneling-ionization
\cite{Keldysh00, Keldysh01, Keldysh02}. In the low-frequency regime
where $\gamma$ is comparable with (or smaller than) one, the
threshold for the ground state is lower than that for the
highly-excited states, as a result of the larger tunneling rate.
However, in the short pulse limit, the ionization threshold for the
ground state also follows the same scaling relation as that observed
for Rydberg states, despite the much smaller $\gamma$
value\cite{note_add}. This observation suggests that the above
discussed displacement ionization is another important mechanism for
the strong-field ionization in short laser pulses, beyond the
tunneling ionization and the multi-photon
ionization\cite{simple_man00, simple_man01, simple_man02}. We note
that the signature of non-zero displacement effect has been
discussed recently by Ivanov \textit{et al.} using a multi-cycle
extreme-ultraviolet pulse\cite{Ivanov}.

\begin{figure}[t]
\centering
  \includegraphics[width=220pt]{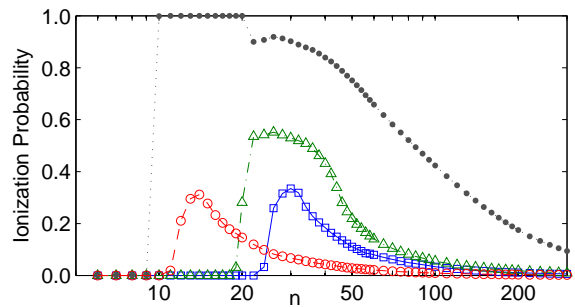}
  \caption{(Color online) ``Ionization window'' for Hydrogen Rydberg atoms initially in a $d$ state.
   The ``symmetric'' pulse in Fig.1(a) is used.
  The applied field amplitudes for the squares, the triangles, the open and the
  solid circles are $2.5kV/cm$, $5kV/cm$, $50kV/cm$ and $100kV/cm$, respectively.
  $t_w=0.1ps$ for the open circles, and $t_w=1ps$ for the others.}
\end{figure}

The proposed displacement-ionization mechanism has several
fundamental differences from both the ionization dynamics induced by
HCP(s) and the conventional field ionization. First of all, the
dominant ionization induced by a short HCP occurs through a sudden
impulsive kick, where a certain amount of momentum is transferred to
an electron\cite{half00, half01, half_add}, and which mainly changes
the electron kinetic energy. In our present situation, however,
displacement ionization is the dominant path, and the ionization
occurs through a sudden spatial displacement of electron, which
mainly changes the potential energy between electron and atomic
nucleus. More importantly, the field ionization threshold in the
short HCP limit has been proved to scale as $n^{-1}$\cite{half00},
which suggests that the more-weakly bound electron is still easier
to be ionized. In contrast, the increasing threshold behavior
($n^2$) induced by the displacement ionization suggests that the
more-deeply bound electron is easier to get ionized. The following
``ionization window'' phenomenon is one benefit from this
displacement-ionization mechanism induced by a single-cycle pulse.
What is more, HCP is a particularly-tailored pulse in practice. A
real optical pulse must satisfy a zero net-force condition ($\int
F(t)dt=0$) which is required by the Maxwell's equations.
Nevertheless, the net spatial displacement ($\int A(t)dt$) can be
nonzero. Therefore, the single-cycle pulse is a natural limit of the
fast-developing short-pulse generation technique. When the electric
field of HCP or the present single-cycle pulse varies slowly enough,
the situation will correspond to the conventional field
ionization\cite{ramp}, which we have discussed in the above context
associated with Fig.1(c) where the differences between the threshold
behavior in the low-frequency limit and that in the short-pulse
limit can be observed clearly. It is these differences that make the
following ``ionization window'' effect predictable and observable.

Based on the above observed threshold behaviors, an interesting
phenomenon can be expected if a specific single-cycle pulse is
applied to ionize a series of Rydberg states. It is shown in Fig.4
for hydrogen $d$ states by using 3D CTMC simulations. To avoid the
possible influence of other $m$ states in the larger ionization
probability, the initial $z$-component of classical angular momentum
is restricted to be from $-0.5$ to $0.5$\cite{MC00}. For the
low-lying Rydberg states where $t_w$ is much longer than $T_{Ryd}$,
the adiabatic ionization is dominant, and the more deeply bound
states cannot be ionized because $F_{th}\varpropto n^{-4}$. However,
for the high-lying Rydberg states where $t_w$ is much shorter than
$T_{Ryd}$, the displacement ionization is dominant, and the more
weakly bound states can be hardly ionized since $F_{th}\varpropto
n^2$. Consequently, an ``ionization window'' is formed by the
different threshold scaling relations between the
adiabatic-ionization regime and the displacement-ionization regime.

Only the Rydberg states falling in the ``ionization window'' are
strongly ionized. The location of this ``window'' is determined by
the pulse duration and can be estimated from Fig.1(c) and Eq.(6).
Its width and height can be adjusted by the field amplitude. The
width can also be estimated using Eq.(6). These features are
demonstrated in Fig.4 by the squares, the open circles and the
triangles, which supplies a possible application in future
experiments as we discussed above. It is interesting to note that
the ionization seems saturated suddenly for the Rydberg states with
$n=10\thicksim20$ in Fig.4 when $F_m=100kV/cm$. To understand this,
we first note that the scaled frequency $\omega n^3=0.3$ for $n=20$
when $t_w=1ps$. For the Rydberg states with $n$ lower than $20$, the
value of $\omega n^3$ is less than $0.3$, and the applied pulse
falls in the low-frequency regime in Fig.4, where the time effect is
much smaller, and the ionization probability can increase quickly
once the applied field strength becomes slightly larger than the
required threshold amplitude, which accounts for the saturated
ionization in Fig.4.

In conclusion, inspired by a recent experiment on the ionization of
sodium Rydberg atom by an intense single-cycle pulse \cite{Bob}, we
have investigated the threshold behaviors for both hydrogen and
sodium atoms in a short single-cycle pulse field. Besides the newly
reported threshold behavior\cite{Bob}, the required threshold field
amplitude was found to scale as $n^2$ in the short pulse limit. This
result has been confirmed by both 3D CTMC simulations and
numerically solving TDS equation. The ionization threshold for the
hydrogen ground state also follows the same scaling behavior in the
short pulse limit. The dominant ionization mechanism was identified
as a sudden displacement of the electron by a short single-cycle
pulse, and an empirical expression was also obtained. This
displacement ionization is a new mechanism for the strong-field
ionization of atoms and molecules, which holds important
implications for the future experiments, especially with short laser
pulses. Finally, an ``ionization window'' was predicted for the
ionization of Rydberg states, which may have potential applications.

We thank Prof. C. H. Greene for the helpful discussions. This work
was supported by the U.S. Department of Energy, Office of Science,
Basic Energy Sciences, under Award number DE-SC0012193.


\begin{references}

\bibitem{external_field}
P. Schmelcher and W. Schweizer (Eds.), \emph{Atoms and Molecules in
Strong External Fields}, (Plenum Press, New York 1998).

\bibitem{intense_field}
T. Brabec (Ed.), \emph{Strong field laser physics}, (Springer, New
York 2008).

\bibitem{Gallagher}
T. F. Gallagher, \emph{Rydberg Atoms}, (Cambridge University Press,
Cambridge, England, 1994), 1st ed..

\bibitem{Gutzwiller}
M. C. Gutzwiller, \emph{Chaos in Classical and Quantum Mechanics},
(Springer, New York, 1990).

\bibitem{COT}
M. L. Du and J. B. Delos, Phys. Rev. Lett. {\bf 58}, 1731 (1987); D.
Kleppner and J. B. Delos, Found. Phys. {\bf 31}, 593 (2001) and
references therein.

\bibitem{simple_man00}
H. B. van Linden van den Heuvell and H. G. Muller, in
\emph{Multiphoton Processes}, edited by S. J. Smith and P. L. Knight
(Cambridge University Press, Cambridge, England, 1988).

\bibitem{simple_man01}
T. F. Gallagher, Phys. Rev. Lett. {\bf 61}, 2304, (1988); E. S.
Shuman, R. R. Jones, and T. F. Gallagher, Phys. Rev. Lett. {\bf
101}, 263001, (2008).

\bibitem{simple_man02}
P. B. Corkum, N. H. Burnett, and F. Brunel, Phys. Rev. Lett. {\bf
62}, 1259, (1989); P. B. Corkum, Phys. Rev. Lett. {\bf 71}, 1994,
(1993).

\bibitem{ramp}
T. W. Ducas, M. G. Littman, R. G. Freeman, and D. L. Kleppner, Phys.
Rev. Lett. {\bf 35}, 366 (1975); T. H. Jeys \textit{et al.}, Phys.
Rev. Lett. {\bf 44}, 390 (1980); J. L. Dexter and T. F. Gallagher,
Phys. Rev. A {\bf 35}, 1934 (1987). G. M. Lankhuijzen and L. D.
Noordam, Adv. At. Mol. Opt. Phys. {\bf 38}, 121 (1997) and the
references therein.

\bibitem{micro_wave}
P. Pillet, W. W. Smith, R. Kachru, N. H. Tran, and T. F. Gallagher,
Phys. Rev. Lett. {\bf 50}, 1042 (1983); P. Pillet, H. B. van Linden
van den Heuvell, W. W. Smith, R. Kachru, N. H. Tran, and T. F.
Gallagher, Phys. Rev. A {\bf 30}, 280 (1984); L. Perotti, Phys. Rev.
A {\bf 73}, 053405 (2006).

\bibitem{localization}
S. Fishman, D. R. Grempel, and R. E. Prange, Phys. Rev. Lett. {\bf
49}, 509 (1982); H. Maeda and T. F. Gallagher, Phys. Rev. Lett. {\bf
93}, 193002 (2004); A. Schelle, D. Delande, and A. Buchleitner,
Phys. Rev. Lett. {\bf 102}, 183001 (2009).

\bibitem{half00}
R. R. Jones, D. You, and P. H. Bucksbaum, Phys. Rev. Lett. {\bf 70},
1236 (1993); C. O. Reinhold, M. Melles, and J. Burgd\"{o}rfer, Phys.
Rev. Lett. {\bf 70}, 4026 (1993); M. T. Frey, F. B. Dunning, C. O.
Reinhold, and J. Burgd\"{o}rfer, Phys. Rev. A {\bf 53}, 2929 (1996).

\bibitem{half01}
C. Raman, C. W. S. Conover, C. I. Sukenik, and P. H. Bucksbaum,
Phys. Rev. Lett. {\bf 76}, 2436 (1996); R. B. Vrijen, G. M.
Lankhuijzen, and L. D. Noordam, Phys. Rev. Lett. {\bf 79}, 617
(1997).

\bibitem{half_add}
A. Emmanouilidou and T. Uzer, Phys. Rev. A {\bf 77}, 063416 (2008);
J. S. Briggs and D. Dimitrovski, New J. Phys. {\bf 10}, 025013
(2008).

\bibitem{Bob}
Sha Li and R. R. Jones, Phys. Rev. Lett. {\bf 112}, 143006 (2014).

\bibitem{FT}
One can show that a fourier-transformed spectra of the ``symmetric''
pulse in Fig.1(a) has a peak at about $\pi/(2.2t_w)$.

\bibitem{MC00}
F. Robicheaux, Phys. Rev. A {\bf 56}, 3358(R) (1997).

\bibitem{Topcu}
T. Top\c{c}u and F. Robicheaux, J. Phys. B {\bf 40}, 1925, (2007).

\bibitem{scaling}
B. Kaulakys, V. Gontis, G. Hermann, and A. Scharmann, Phys. Lett. A
{\bf 159}, 261, (1991).

\bibitem{note}
To get exactly the same results, the initial angular momentum
$\mathbf{L}$ also needs to be scaled as $\mathbf{L}/n$ accordingly,
but the influence of this angular-momentum term in the Hamiltonian
is relatively small and is hardly to be observed clearly, especially
for the Rydberg states.

\bibitem{SFI}
F. Robicheaux, C. Wesdorp, and L. D. Noordam, Phys. Rev. A {\bf 62},
043404 (2000) and the references therein.

\bibitem{TDS}
F. Robicheaux, J. Phys. B {\bf 45}, 135007, (2012).

\bibitem{displacement}
C. Wesdorp, F. Robicheaux, and L. D. Noordam, Phys. Rev. Lett. {\bf
87}, 083001 (2001).

\bibitem{book}
C. E. Burkhardt and J. J. Leventhal, \emph{Topics in Atomic
Physics}, (New York, Springer, 2006).

\bibitem{Keldysh00}
L. V. Keldysh, Sov. Phys. JETP {\bf 20}, 1307 (1965).

\bibitem{Keldysh01}
V. S. Popov, Phys. Usp. {\bf 47}, 855 (2004) and the references
therein.

\bibitem{Keldysh02}
T. Top\c{c}u and F. Robicheaux, Phys. Rev. A {\bf 86}, 053407,
(2012).

\bibitem{note_add}
There is a small contribution from tunneling ionization, which can
decrease the corresponding threshold value slightly, but it can be
hardly seen in Fig.3(b).

\bibitem{Ivanov}
I. A. Ivanov, A. S. Kheifets, K. Bartschat, J. Emmons, S. M. Buczek,
E. V. Gryzlova, and A. N. Grum-Grzhimailo, Phys. Rev. A {\bf 90},
043401, (2014).


\end{references}
\end{document}